\documentclass[10pt,conference]{IEEEtran}
\IEEEoverridecommandlockouts

\usepackage{siunitx}
\usepackage{cite}
\usepackage{amsmath,amssymb,amsfonts, amsthm}
\usepackage{tikz}
\usetikzlibrary{automata, positioning, arrows,calc}
\usetikzlibrary{quantikz}
\usepackage{algorithmic}
\usepackage{graphicx}
\usepackage{textcomp}
\usepackage{xcolor}
\usepackage{caption} 
\usepackage{subcaption}

\usepackage{hyperref}
\usepackage{mathtools}
\usepackage{multirow}
\usepackage{graphicx}
\usepackage{multicol}
\usepackage{array}
\usepackage{enumitem}

\newtheorem{cor*}{Corollary}

\newcommand{\mpfont}{\scriptsize}

\ifx\noeditingmarks\undefined
    \newcommand{\MPworker}[2]{{\color{#1}\vrule\vrule}{\marginpar{\color{#1}\mpfont #2}}}
\else
    \newcommand{\MPworker}[2]{}
\fi

\begin{document}

\title{
Optimization of C-band quantum traffic coexisting with O-band classical traffic: preliminary results
}

\makeatletter
\newcommand{\linebreakand}{%
  \end{@IEEEauthorhalign}
  \hfill\mbox{}\par
  \mbox{}\hfill\begin{@IEEEauthorhalign}
}
\makeatother

\author{
\IEEEauthorblockN{Laura d'Avossa ~\IEEEmembership{Graduate Student~Member,~IEEE}, Elena Montella, Marco Grillo, \\
Angela Sara Cacciapuoti ~\IEEEmembership{Senior~Member,~IEEE}, Marcello Caleffi~\IEEEmembership{Senior~Member,~IEEE} }

\thanks{
The authors are with the www.QuantumInternet.it research group, University of Naples Federico II, Naples, 80125 Italy.
This work has been funded by the European Union under Horizon Europe ERC-CoG grant QNattyNet, n.101169850. Views and opinions expressed are however those of the author(s) only and do not necessarily reflect those of the European Union or the European Research Council Executive Agency. Neither the European Union nor the granting authority can be held responsible for them.}

}

\maketitle

\begin{abstract}
The coexistence of quantum and classical signals in the same optical fiber is a critical challenge for the deployment of quantum networks. Indeed, selecting an optimal channel for quantum signal transmission is crucial to minimize noise arising from co-propagating classical signals. This work experimentally investigates spontaneous Raman scattering (SpRS), a major source of noise in signals transmitted along the same fiber. Unlike most previous studies relying on narrow-linewidth laboratory lasers or architectures based on spatial or temporal multiplexing of quantum and classical signals, we employ commercial SFP optical transceivers and standard single-core single-mode fiber for the transmission of quantum and classical signals in the same fiber, reflecting conditions typical of deployed urban fiber infrastructures.
Building on these measurements, we derive a compact and predictive model that captures the Raman scattering profile, enabling accurate estimation of SpRS noise as a function of source power, wavelength, and fiber length. A key outcome of this work is that the proposed model is independent of the specific optical source used, demonstrating its generality and robustness. The model can therefore be used for the identification of optimal C-band channels for quantum signal allocation, namely those least affected by SpRS noise generated by co-propagating O-band classical traffic.
These results pave the way for a parameter-robust description of Raman scattering applicable to diverse fiber-based systems.
\end{abstract}


\section{Introduction}
\label{sec:1}

\begin{figure} 
        \centering
        \includegraphics[width=\linewidth]{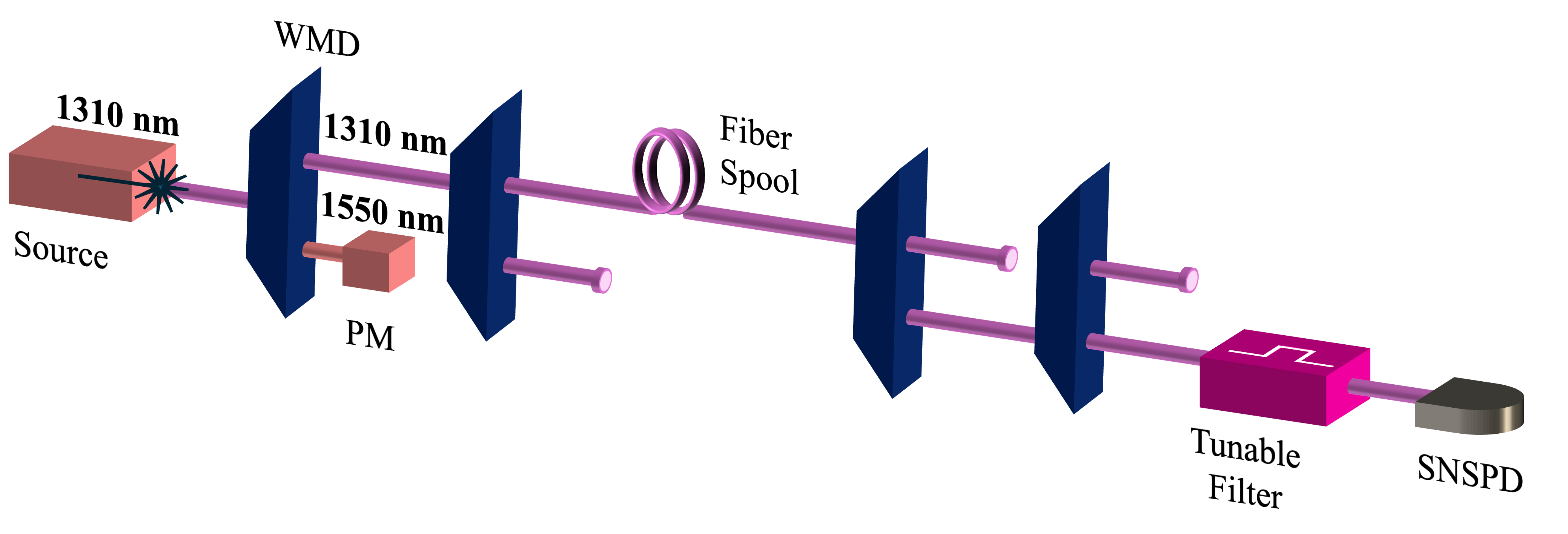}
        \caption{Experimental setup for the measurement C-band SpRS photons generated by an O-band classical signal. A classical source at 1310 nm is spectrally filtered using cascaded WDM to suppress out-of-band components before propagation through optical fiber spools of different lengths. To monitor source stability, the common port C of the first WDM in the cascade is connected to a power meter (PM). At the receiver side, a second WDM cascade isolates the C-band spectral region. A narrowband tunable filter is used to scan the spectrum across ITU channels 15 to 63. The resulting SpRS photons are detected with a superconducting nanowire single-photon detector (SNSPD).}
        \hrulefill

           \label{fig:01}
\end{figure}\hfill

Entanglement distribution is the key enabling functionality for future quantum networks, enabling protocols and application with no counterpart in classical networks, such as quantum key distribution, quantum teleportation, and distributed quantum computing \cite{CacCalTaf-19}. Optical photons constitute the ideal candidates as entanglement carriers \cite{CalDavHan-25, DavCacCal-25}, and are thus widely exploited for long-distance quantum communication over optical fiber links. 
While preliminary demonstrations of entanglement distribution have relied on dedicated infrastructure employing dark fibers \cite{CraLazPorSekFlaNam-24}, the large-scale deployment of quantum networks requires leveraging the existing telecommunications fiber infrastructure. Indeed, deploying infrastructure exclusively dedicated to quantum traffic would be impractical and prohibitively expensive at large scale.
Nonetheless, multiplexing quantum signals with higher-power classical traffic can introduce significant noise. Here, one of the major impairments arises from Spontaneous Raman Scattering (SpRS), a process whereby photons belonging to the classical optical signal are scattered over a broad spectrum and can fall within the spectral band of the co-propagating quantum signal \cite{FerXavTemVon-14}.

Within the \textit{National Quantum Internet.it testbed} deployed at \textit{University of Naples Federico II} \cite{CalDavFlo-26, IlMattino-26}, the quantum signal -- i.e., the entanglement carriers --  can be generated across the whole \textit{C-band}, namely, across any of forty DWDM\footnote{DWDM is a  multiplexer/demultiplexer technique that divides the C-band into 40 channels, each with a spectral spacing of 100 GHz.}
channels spanning the $1520$-$1577$nm portion of the optical spectrum. The rationale for this choice is that this portion of the spectrum exhibits the lowest attenuation in standard single-mode optical fibers. Indeed, the attenuation is approximately $~0.2$ dB/Km, making the C-band ideal for long-distance quantum communications \cite{ClaYanWanNejSimJos-24}. Clearly, 
due to the huge power disparity between quantum and classical signals -- up to twelve orders of magnitude in our testbed -- classical traffic cannot be co-propagated within the same spectral band. For this reason, in our testbed classical signals are placed in the \textit{O-band}, the spectral region spanning approximately $1260$–$1360$ nm, which is commonly used in conventional optical networks for short-to-medium links.

 \begin{figure} 
        \centering
        \includegraphics[width=\linewidth]{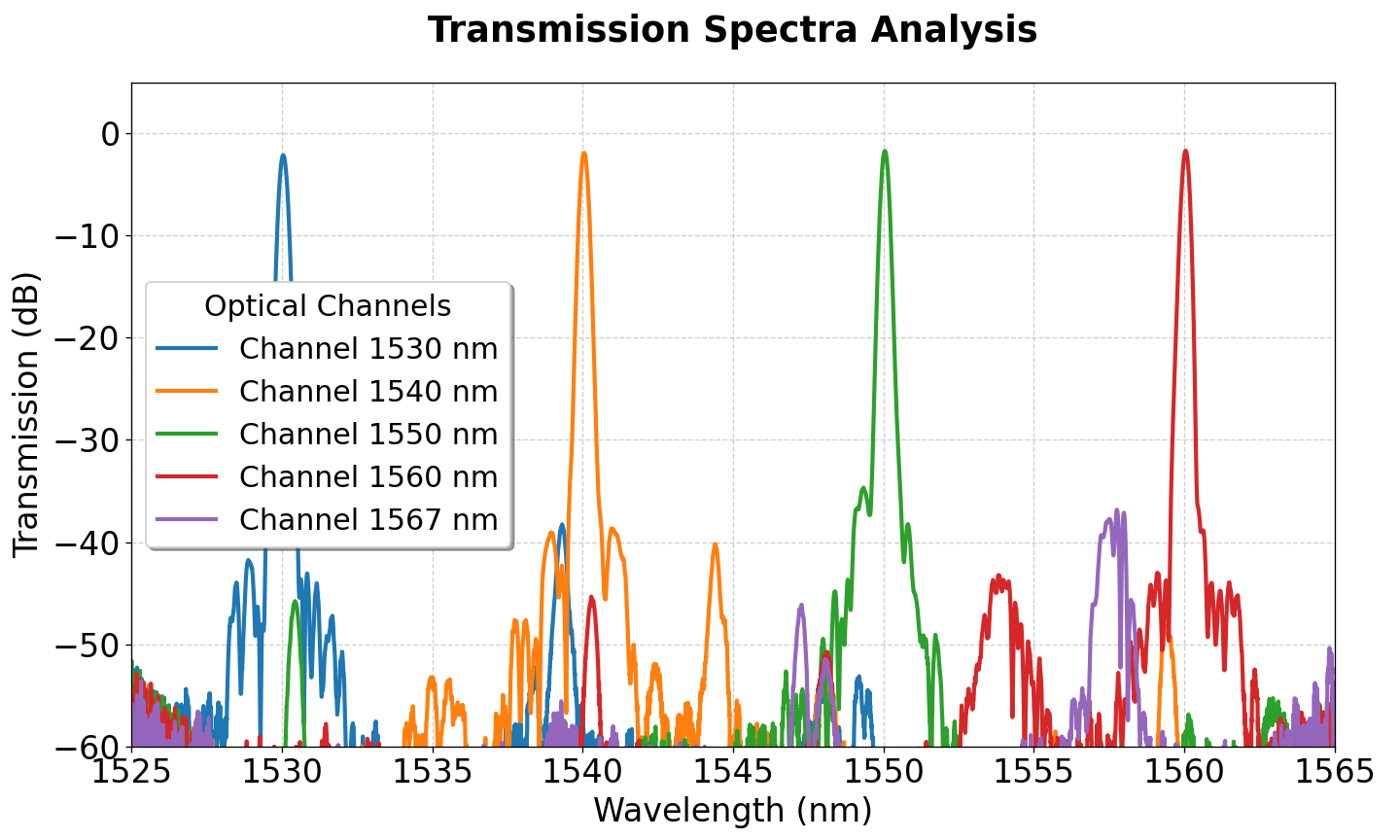}
        \caption{Transmission profile of the tunable narrowband filter. The measured response exhibits a sharp roll-off at the edges of the nominal passband of each selected channel. The filter achieves an out-of-band suppression of approximately -40 dB even near the passband edges, yielding a sharply defined passband response and ensuring high channel selectivity.
        }
        \hrulefill

           \label{fig:filter}
    \end{figure}\hfill

\begin{figure*}[t] \centering \begin{subfigure}{0.48\textwidth} \centering \includegraphics[width=\linewidth]{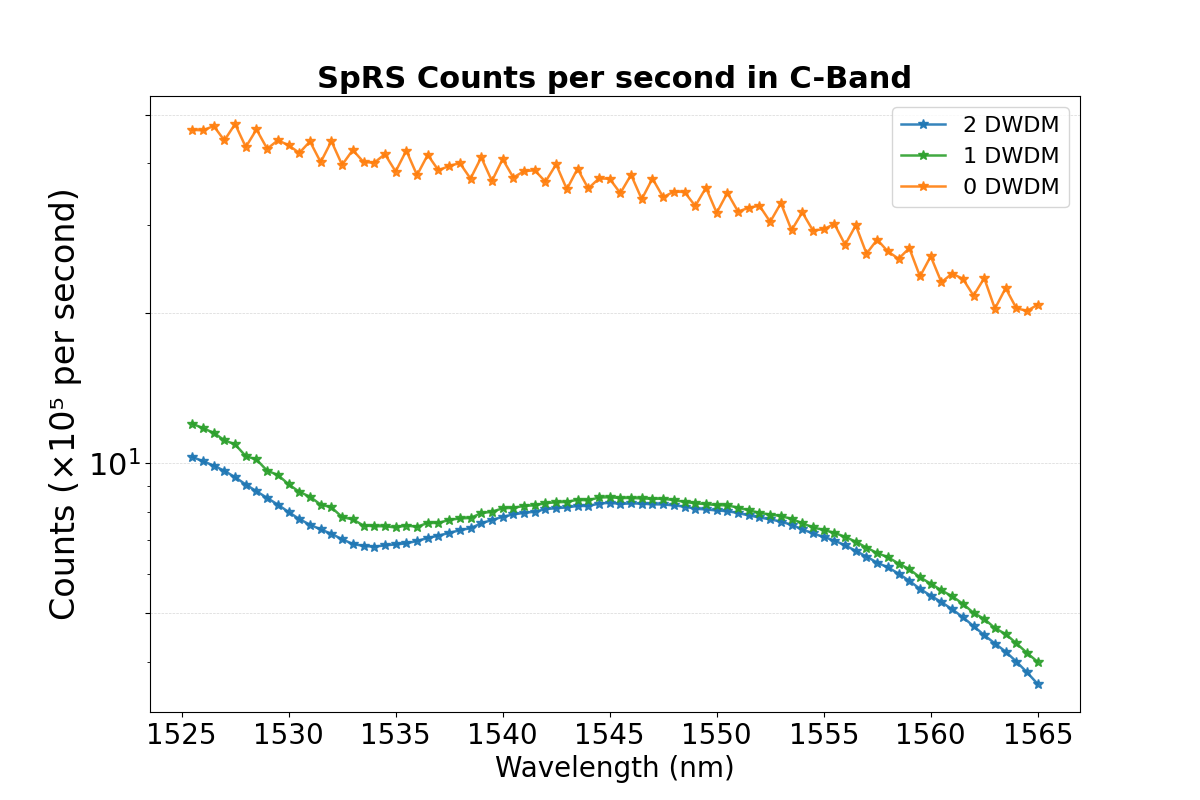} \caption{Variation of WDM components to isolate the O-band signal, with a fixed number of DWDMs for isolating the signal in the C-band.} 
\label{fig:02.1} 
\end{subfigure} 
\hfill 
\begin{subfigure}{0.48\textwidth} \centering \includegraphics[width=\linewidth]{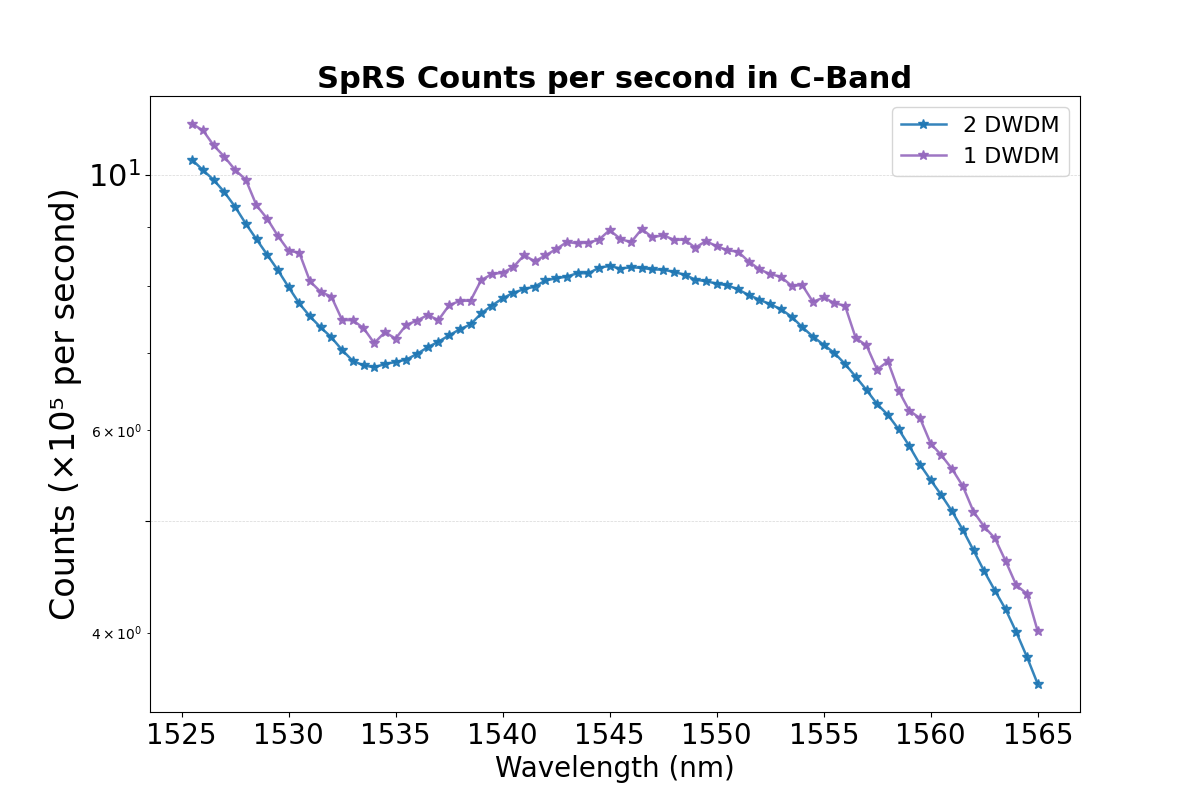} 
\caption{Variation of WDM components to isolate the C-band signal, with a fixed number of WDMs for isolating the signal in the O-band.} 

\label{fig:02.2} 
\end{subfigure} 
\caption{SpRS counts $C_{SpRS}$ generated in the C-band from a classical O-band source at 1310 nm.} 
\hrulefill
\label{fig:02} 
\end{figure*}

In this context, identifying the optimal quantum channel -- i.e., the DWDM channel in the C-band experiencing the lowest SpRS induced by co-propagating classical signals in the O-band -- is a key preliminary task for enabling the coexistence of quantum and classical traffic within the same fiber.

Although theoretical models describing spontaneous Raman scattering in silica fibers are well-established \cite{HolCan-02}, they typically provide a generalized spectral description and fail to fully capture the fine structure observed in standard single-mode fibers. Additionally, most studies on quantum-classical coexistence rely on architectures that mitigate nonlinear interactions through spatial or temporal separation, such as multicore fiber systems that exploit spatial multiplexing \cite{WuRbDiS-25} or temporally interleaved transmission schemes that separate quantum and classical channels in time \cite{WanRolBri-24}.
In contrast, this work considers a co-propagation scenario in a standard single-core single-mode fiber. 
Accordingly, classical and quantum signals propagate simultaneously in the same physical medium, without any spatial or temporal multiplexing. 
Indeed, this is a preliminary yet fundamental analysis to assess where the amount of Raman scattering generated by the classical signal is "tolerable" by the quantum signal.
To this end, we experimentally study C-band SpRS noise generated by O-band classical traffic, using commercial SFP optical transceivers. This choice allows us to capture realistic spectral characteristics of the source, which can significantly influence the Raman noise profile in practical implementations. The analysis is further extended to different fiber lengths, up to 5 km, representative of urban scenarios.

It is also worth noting that previous experimental investigations in similar testbeds, where quantum and classical traffic coexisted in the same fiber, typically considered inverse wavelength allocation, with quantum signals in O-band and classical traffic in C-band \cite{TalHesDav-26, TalHesTho-26}, whereas this work focuses on the opposite coexistence scenario.

Based on these experimental results, we develop a compact and predictive model that aims at capturing the observed Raman spectral variations. The model allows us to estimate SpRS-induced noise as a function of classical launch power, fiber length, and wavelength, and provides a practical tool for designing and optimizing quantum-classical coexistence in fiber optic systems.


\section{Raman Scattering Model}

Spontaneous Raman scattering (SpRS) constitutes one of the primary mechanisms through which photons propagating within a certain portion of the spectrum can be scattered from one band to another. 
In our case study, energy is transferred from the original O-band photons used for classical traffic, generating Stokes and anti-Stokes components that extend into the C-band and may interfere with co-propagating channels \cite{EraWal-10,HolCan-02}.
The SpRS power generated is \cite{FroDynLuc-15,ThoKanKum-25, ThoKanXie-23}:
\begin{align}
    P_{Raman}(\lambda_o\lambda_c)= \beta(\lambda_o\lambda_c)P_o\Delta_\lambda L_{eff}
    \label{eq:01}
\end{align}
where $P_o$ is the original power launched, $\Delta \lambda$ is the full width at half maximum bandwidth of the filter applied to the received photons.
$L_{eff}$ is the effective interaction length given by \cite{ThoKanKum-25, FroDynLucSha-15}:

\begin{align}
    L_{\rm eff}= \frac{e^{-\alpha_c L} - e^{-\alpha_o L}}{\alpha_o - \alpha_c}
    \label{eq:02}
\end{align}
where $\alpha$ is the fiber attenuation coefficient with the subscripts $o$ and $c$ referring to the O-band and C-band, respectively, and $L$ is the physical length of the fiber.

Finally, $\beta(\lambda_o, \lambda_c)$ is the Raman scattering coefficient, which quantifies the energy transferred between spectral components, corresponding to the frequency shift between wavelengths $\lambda_o$ (O-band) and $\lambda_c$ (C-band).

\begin{figure*}[t]
    \centering

    \begin{subfigure}{0.48\textwidth}
        \centering
        \includegraphics[width=\linewidth]{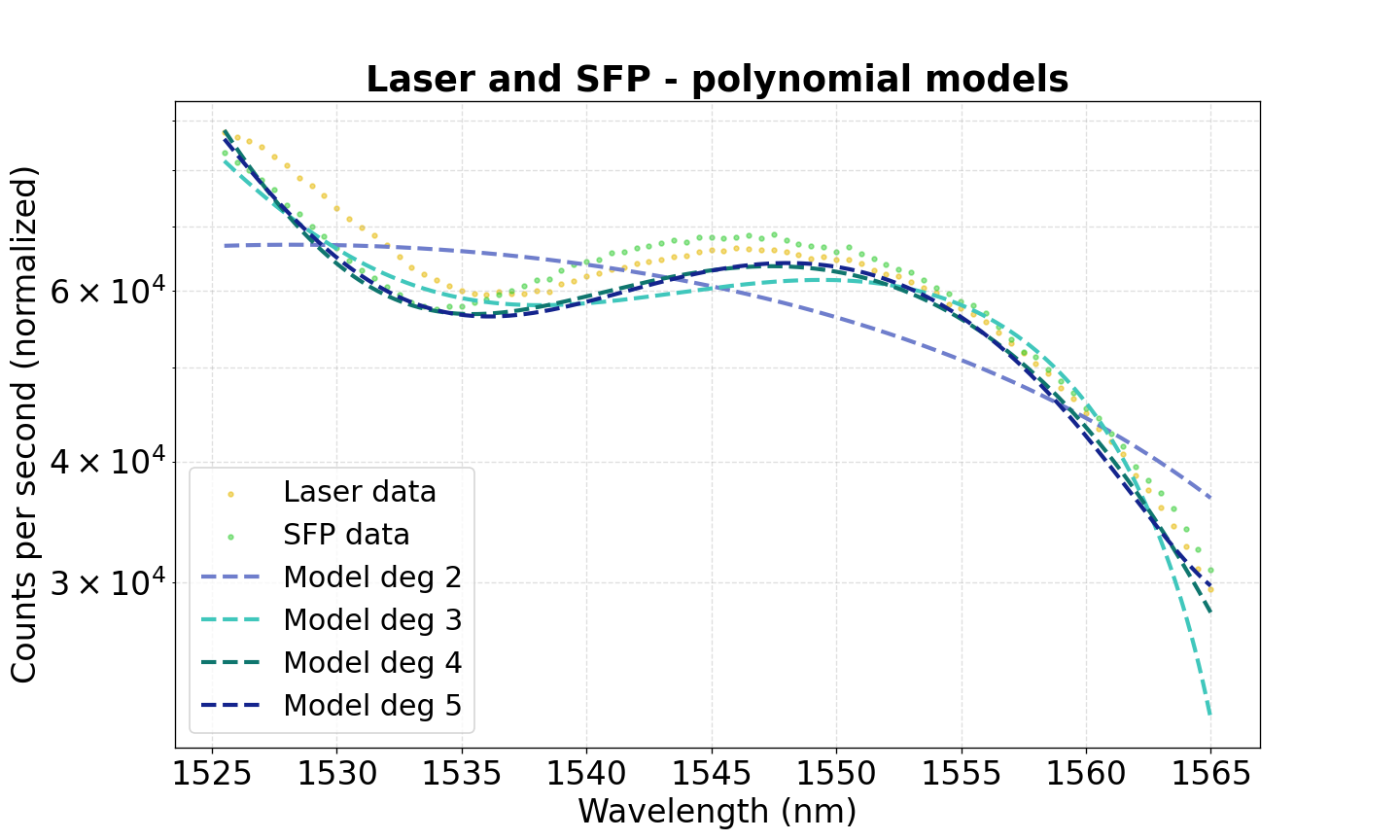}
        \caption{500 m}
        \label{fig:03.1}
    \end{subfigure}
    \hfill
    \begin{subfigure}{0.48\textwidth}
        \centering
        \includegraphics[width=\linewidth]{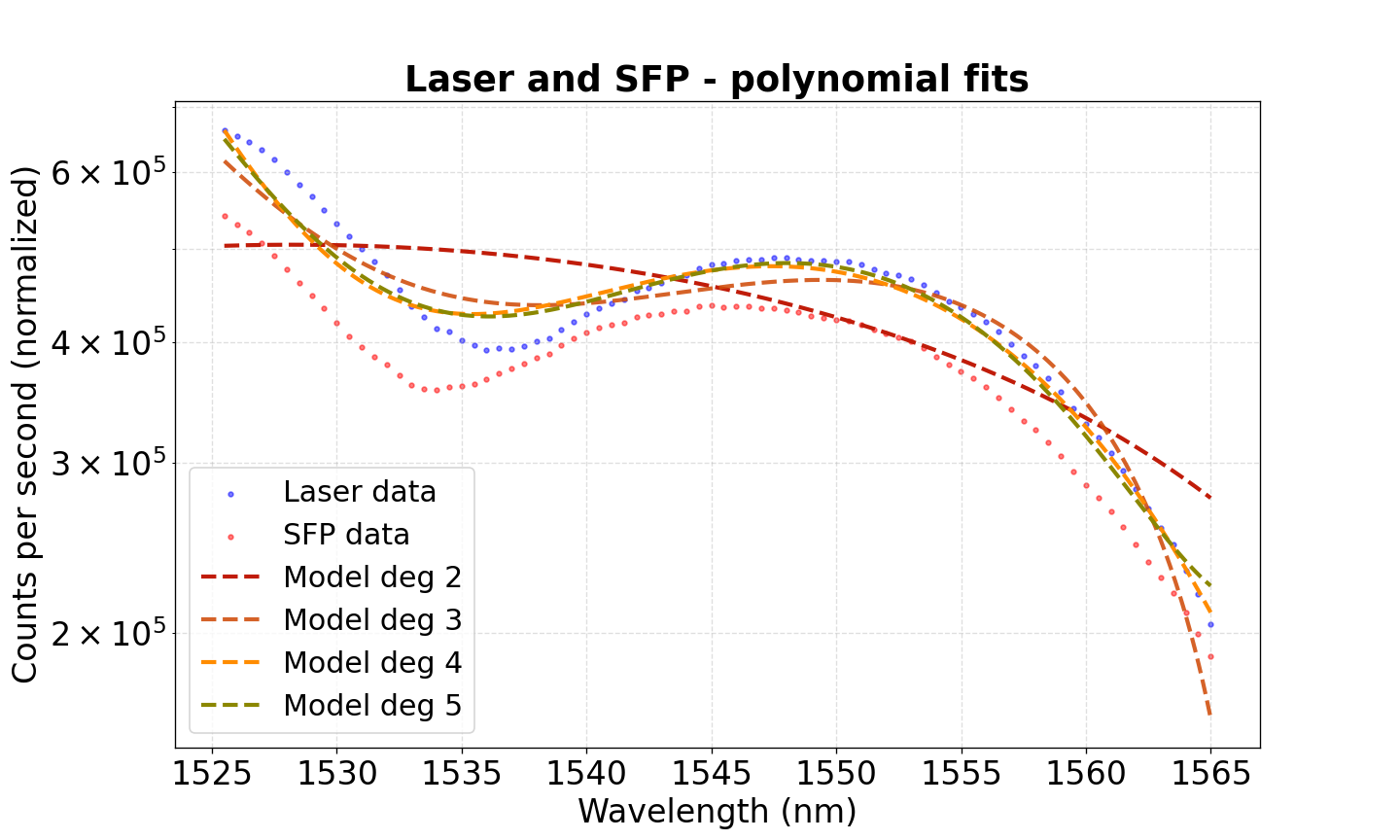}
        \caption{5000 m}
        \label{fig:03.2}
    \end{subfigure}

    \caption{SpRS counts $C_{SpRS}$ for different fiber lengths, comparing experimental measurements with theoretical model of Eq.~\ref{eq:04}. The model is evaluated using polynomial fit of increasing order for $A(\lambda)$. 
    }
    \label{fig:03}
    \hrulefill

\end{figure*}

\begin{figure}[t]
    \centering
    \includegraphics[width=\linewidth]{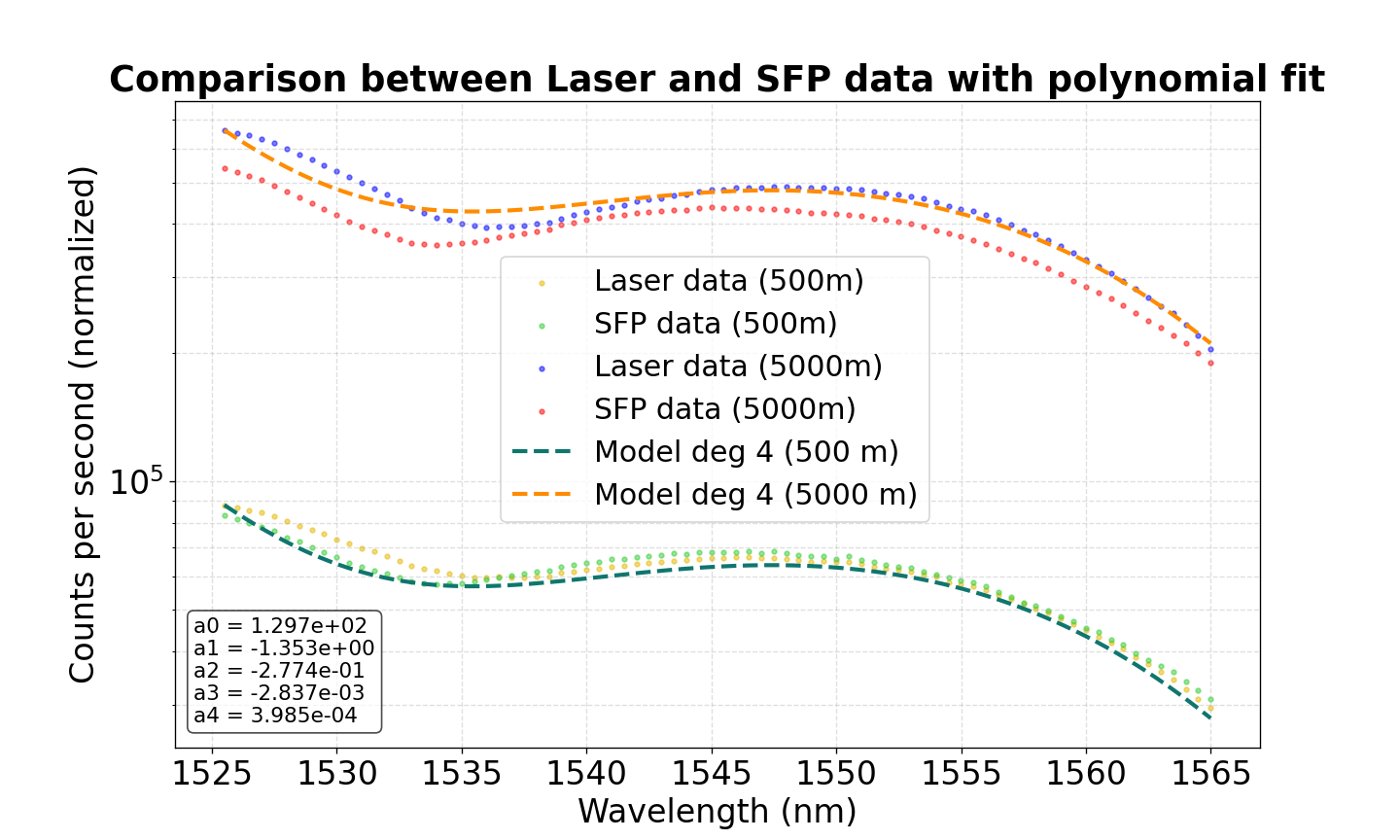}
    \caption{Comparison between measured $C_{SpRS}$, acquired using two optical sources and fiber spool lengths of 500 m and 5 km, with the corresponding theoretical model. The experimental data are normalized to account for the different output powers of the sources. The theoretical curves are obtained from the proposed $C_{SpRS}$ model using a fourth-degree polynomial approximation for $A(\lambda)$, enabling a direct validation of the fitting performance against the measurements.}
    \hrulefill

    \label{fig:04}
\end{figure}

\section{Experimental setup}




    
    
    

The experiment is conducted within the \textit{National Quantum Internet.it testbed} at the Monte Sant’Angelo campus of the University of Naples Federico II. The experimental setup is illustrated in Fig.~\ref{fig:01}.
To evaluate C-band SpRS photons generated by O-band classical traffic, we employ different classical sources with a central wavelength of 1310 nm\footnote{1310 nm is commonly used in classical optical communications, where fiber dispersion is minimal and signal distortion is reduced.}. The source output is filtered through a cascade of two WDM filters to isolate the O-band and suppress any spectral components outside 1310 nm potentially generated by the source. 
To monitor source stability, the common port C of the first WDM in the cascade -- used to isolate the O-band component -- is connected to a power meter. This configuration allows for real-time verification that the power emitted by the source does not undergo significant variations during measurement acquisition.
The O-band signal is then transmitted through optical fiber spools of different lengths, and then filtered by a second WDM cascade configured to isolate the C-band signal.
The C-band signal is filtered through a selective narrowband filter, with a bandwidth of 25 GHz ($200$ $pm$). The scanning procedure is performed with a spacing of 500 pm, covering the range from ITU channels 15 to 63. As can be seen from the transmission profile shown in Fig.~\ref{fig:filter}, the measured response exhibits a sharp roll-off at the edges of the nominal passband of each selected channel. The filter achieves an out-of-band suppression of approximately -40 dB even near the passband edges, yielding a sharply defined passband response and ensuring high channel selectivity.

The generated SpRS photons are measured using a Superconducting Nanowire Single-Photon Detector (SNSPD) operating with a cryogenic system of $2.2\,\mathrm{K}$ and a pressure of approximately $10^{-6}\,\mathrm{mbar}$. In these conditions and with the detector characterization performed by tuning the bias current, at $1550\,\mathrm{nm}$ the system achieves detection efficiency (SDE) of about $80\%$, a timing jitter of $50\,\mathrm{ps}$, and a dark count rate (DCR) of approximately $100\,\mathrm{cps}$.
These high-performance characteristics enable a precise study of the noise generated by scattering processes, ensuring that even weak Raman scattering signals can be accurately measured and distinguished from background noise.


\section{Signals filtering}
\label{sec:Signals filtering}

\begin{figure*}[t]
    \centering

    \begin{subfigure}{0.48\textwidth}
        \centering
        \includegraphics[width=\linewidth]{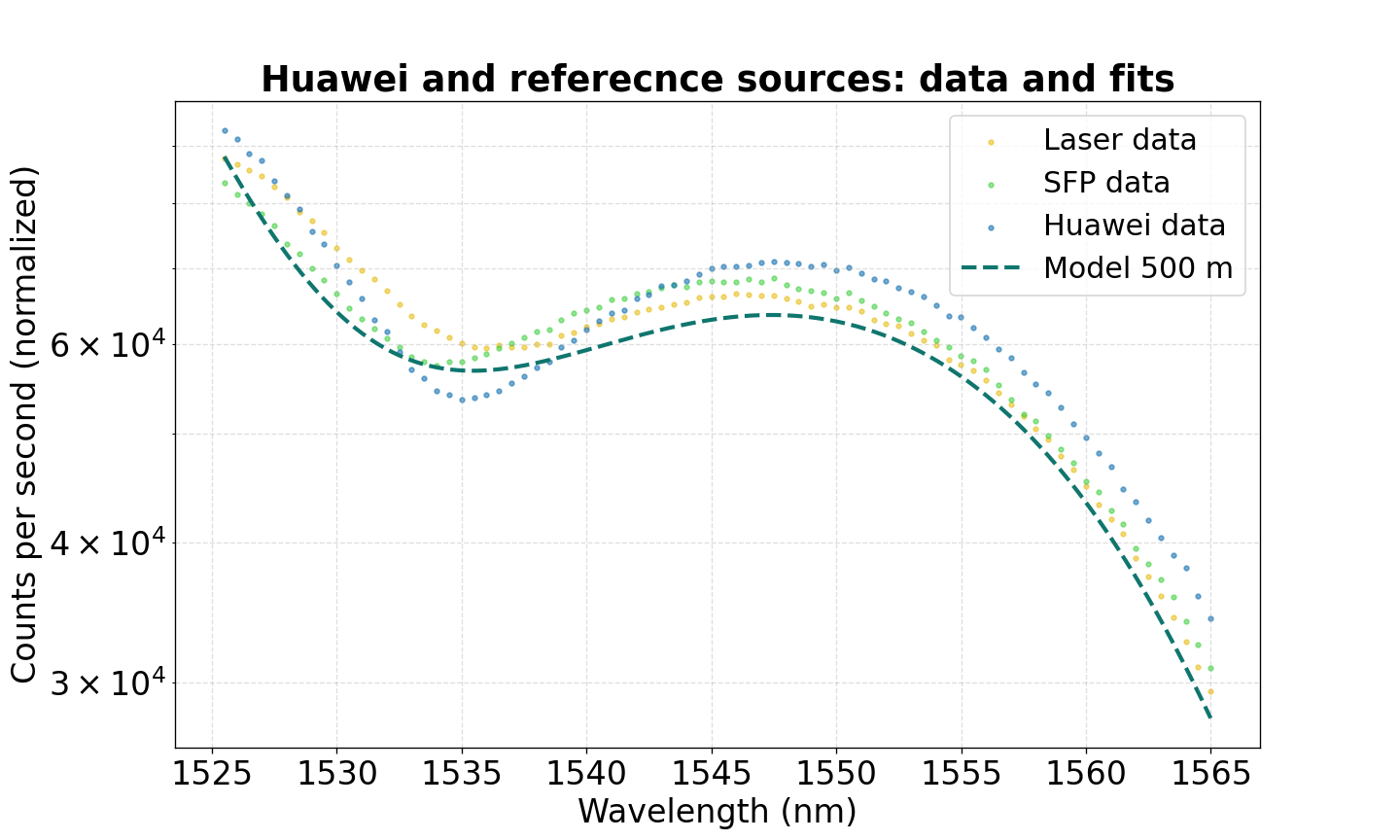}
        \caption{500 m}
        \label{fig:05.1}
    \end{subfigure}
    \hfill
    \begin{subfigure}{0.48\textwidth}
        \centering
        \includegraphics[width=\linewidth]{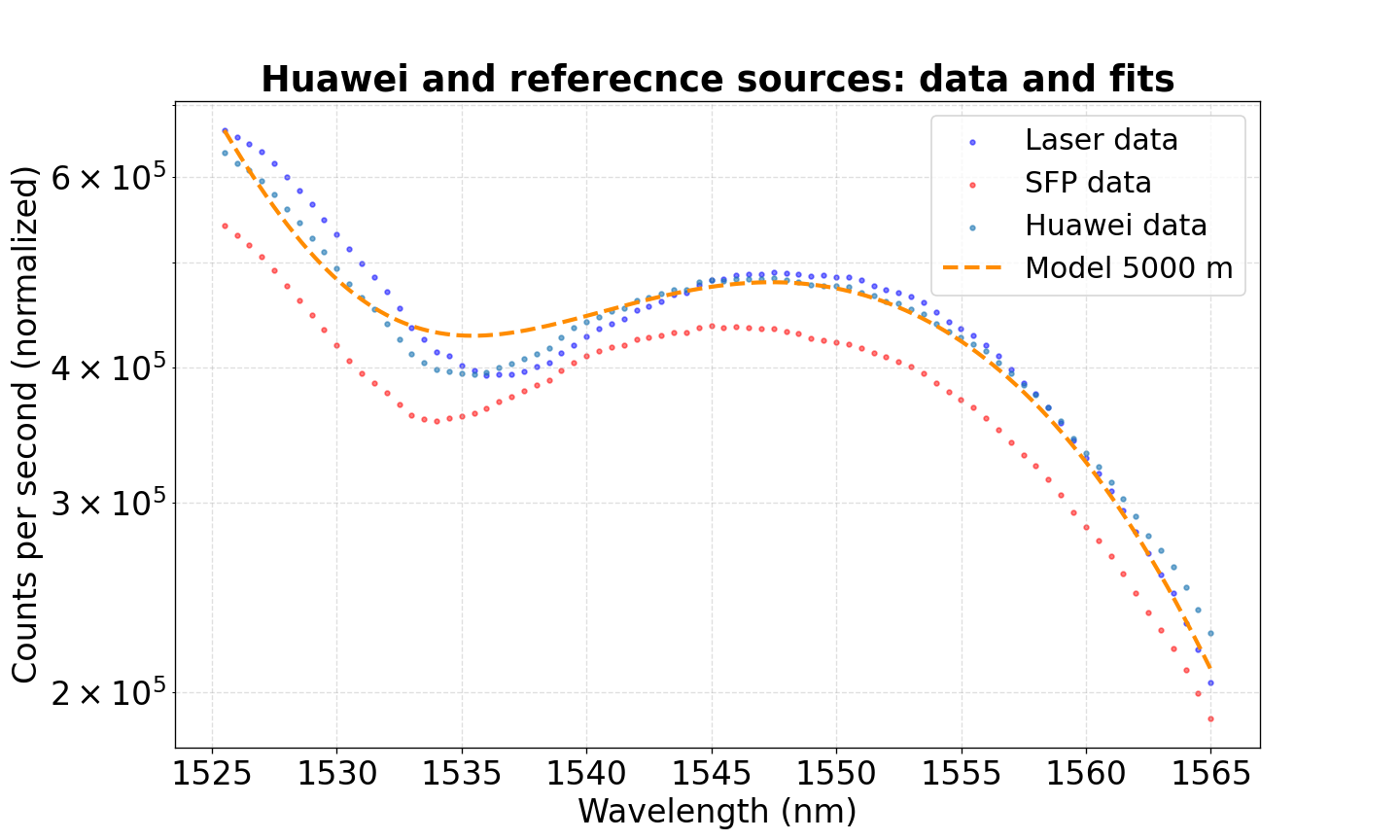}
        \caption{5000 m}
        \label{fig:05.2}
    \end{subfigure}

    \caption{SpRS counts $C_{SpRS}$ for different fiber lengths: comparison between experimental data and derived fits for the third source (Huawei SFP) and the previously analyzed reference classical sources.}
    \label{fig:05}
    \hrulefill

\end{figure*}

The number of WDM components used in the cascades plays a crucial role, as it directly affects signal filtering in both the O-band and the C-band. To investigate this effect, a dedicated study is carried out to assess the impact of the cascaded WDM filters on the traffic.
The experiment is conducted using a commercial SFP source\footnote{An SFP transmitter is a compact optical module that converts electrical signals into optical signals for transmission over optical fiber. In the proposed work, the device used is a Finisar FTLF1321P1BTL module.} with an output power of approximately –1.4 dBm at $\lambda_o = 1310$ nm\footnote{1310 nm is typically used in classical optical communications due to the low dispersion of chromatic fiber in this region, which reduces signal distortion during propagation} over a 5 km fiber spool.

It is evident from Fig.~\ref{fig:02.1} that the presence of at least one WDM isolating the O-band traffic at 1310 nm is essential, namely the count rate decreases considerably when moving from zero to one WDM. This suggests that a large portion of the counts obtained in the absence of a WDM to isolate the 1310 nm signal are actually  photons not spectrally centered at the target wavelength.
The addition of a second WDM that filter O-band photons further smooths the signal, reducing fluctuations. In other words, while a large portion of the external signal at 1310 nm is effectively suppressed, a residual fraction of the signal remains detectable and should be removed with a second stage of WMD.
After the fiber spool, the presence of WDMs is essential to isolate the C-band signal, i.e. the photons generated from the O-band due to the Raman scattering.  Fig.~\ref{fig:02.2} shows that increasing the number of DWDMs in the C-band reduces fluctuations in the measured counts, producing smoother and more uniform curves.

As a result, although increasing the number of DWDM components introduces additional insertion losses and reduces the measured counts, the configuration with the maximum number of DWDMs before and after the fiber is preferred, as it more effectively isolates the signal of interest.

Even from this preliminary analysis, it is already possible to identify a spectral minimum. Indeed, a clear minimum is observed around 1535 nm (channel 44), suggesting that this channel is the most suitable for the propagation of a quantum signal in C-band in co-propagation with the O-band classical signal.


\section{Experimental Results and predicted model}

To characterize the Raman scattering behaviour and derive a predictive model, the measurements are performed using two different O-band sources operating at 1310 nm and different launch powers over fiber spools of 500 m and 5 km.
The first source is a commercial SFP optical transceiver\footnote{The commercial SFP considered in this analysis is the same device employed in the WDM cascade study presented in Sec.~\ref{sec:Signals filtering}, namely the Finisar FTLF1321P1BTL module.}, while the second is a narrow-linewidth laser, which is included as a useful reference.

From the measured SpRS counts $C_{SpRS}$ it is possible to derive the following model:

\begin{align}
    C_{SpRS}(S_{src},\lambda, L) = S_{src} \, A(\lambda) \, \frac{e^{-\alpha_c L} - e^{-\alpha_o L}}{\alpha_o - \alpha_c}
    \label{eq:04}
\end{align}
where $S_{src}$ is a source-dependent scale factor accounting for the different launch powers of the employed sources. 
The dependence on the fiber propagation length is instead entirely described by the multiplicative attenuation term $\frac{e^{-\alpha_c L} - e^{-\alpha_o L}}{\alpha_o - \alpha_c}$, that constitutes the attenuation factor expressed in Eq.\ref{eq:02}. 
Specifically, $\alpha_o$ and $\alpha_c$ are fixed to physically realistic values, $\alpha_o = 0.35$~dB/km for the pump O-band signal and $\alpha_c = 0.2$~dB/km for the Raman scattered photons in the C-band.
$A(\lambda)$ represents the wavelength-dependent Raman spectral profile, namely the Raman scattering coefficient at a fixed $\lambda_o=1310$ as a function of $\lambda_c \in [1525,\, 1565]$, modeled through a polynomial function derived from the experimental measurements obtained with both sources.


We evaluate polynomial fits of varying degrees, in order to identify the one that best approximates the experimental trend while keeping the polynomial degree as low as possible and avoiding any loss of generality. Fig.~\ref{fig:03} shows the counts obtained from the two sources with different polynomial fits\footnote{The figures are shown separately to highlight the differences between polynomial fits of varying degrees, but all the fits are derived using data from both the 500 m and 5 km fiber spools.}.
It is evident that the fourth-degree polynomial provides the best approximation of the experimental trend, while the fifth-degree polynomial does not offer any significant improvement.

Fig.\ref{fig:04} compares the experimental $C_{SpRS}$ counts obtained using the two optical sources for both 500 m and 5 km fiber spools\footnote{The experimental data were normalized to compensate for the different output powers of the optical sources.} with the corresponding theoretical curves derived from the $C_{SpRS}$ model, where $A(\lambda)$ is approximated by a fourth-degree polynomial.

\begin{figure}[t]
    \centering
    \includegraphics[width=\linewidth]{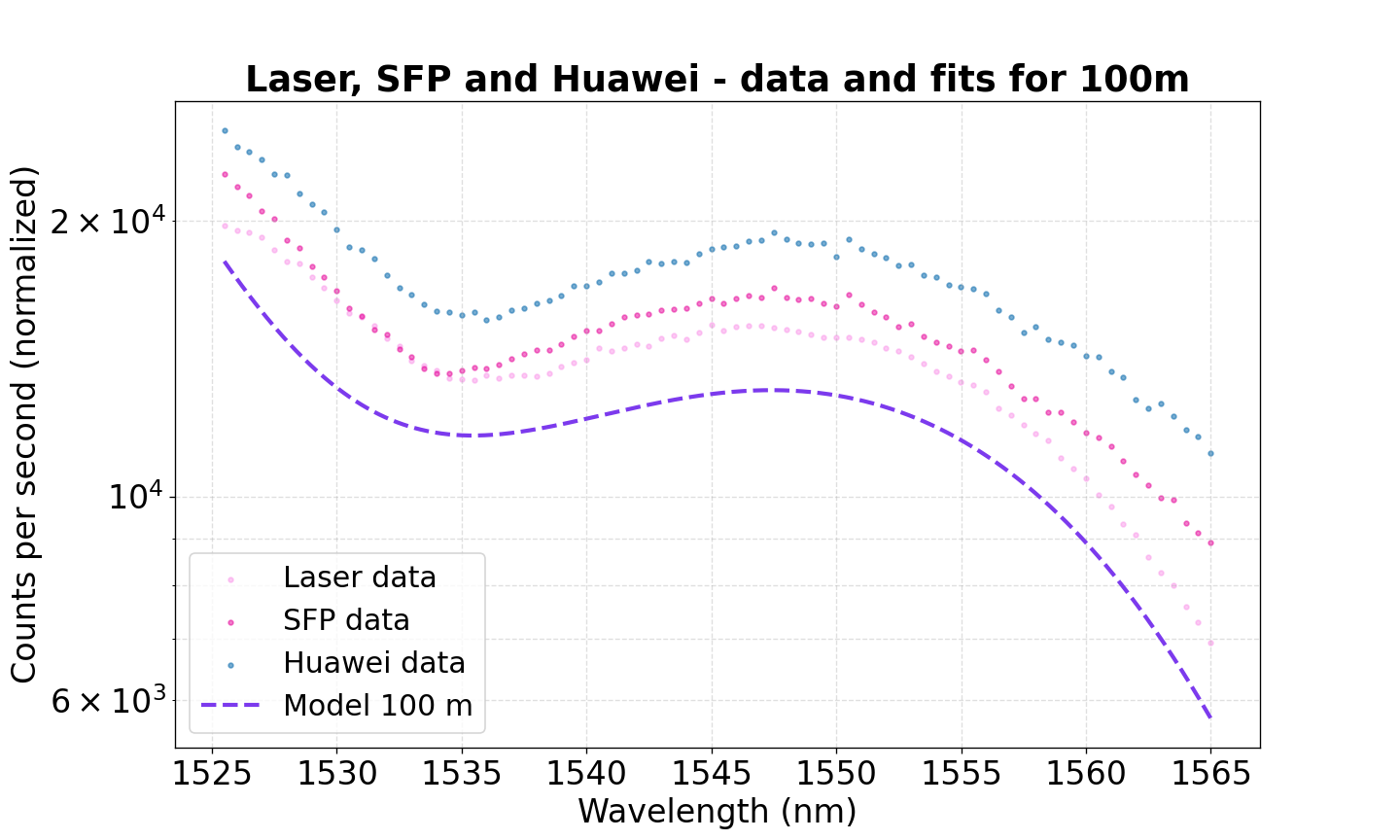}
    \caption{Experimental data and derived fits for all three sources (Laser, Commercial Source, and Huawei) measured on a 100 m fiber spool.}
        \hrulefill

    \label{fig:06}
\end{figure}

\begin{table}[t]
\centering
\renewcommand{\arraystretch}{1.2}
\begin{tabular}{|>{\centering\arraybackslash}m{2.5cm}
                |>{\centering\arraybackslash}m{2.5cm}
                |>{\centering\arraybackslash}m{2.5cm}|}
\hline\hline
\textbf{Source} & \textbf{Fiber length (m)} & \textbf{Relative error $\epsilon(\%)$} \\
\hline\hline
Laser & 100  & 14.9 \\
Laser & 500  & 5.1 \\
Laser & 5000 & 3.7 \\
\hline
Commercial source & 100  & 21.2 \\
Commercial source & 500  & 5.1 \\
Commercial source & 5000 & 13.3 \\
\hline
Huawei & 100  & 32.2 \\
Huawei & 500  & 9.1 \\
Huawei & 5000 & 2.7 \\
\hline\hline
\end{tabular}
\caption{Relative error $\epsilon(\%)$ of the derived fits for different sources and fiber lengths.}
\label{tab:01}
\end{table}




 
To assess the quality of the proposed fit, the relative error $\epsilon(\%)$ between the model and the experimental data is evaluated as follows:

\begin{align}
\epsilon(\%) = \frac{1}{N} \sum_{i=1}^{N} \frac{|C_{SpRS}(\lambda_i) - C_{fit}(\lambda_i)|}{C_{SpRS}(\lambda_i)} \times 100,
\end{align}

where $C_{fit}$ denotes the model predictions and $N$ the total number of experimental measurement points. This approach allows a precise comparison of the spectral and length-dependent behaviour of Raman-induced counts across different sources and fiber lengths.
The relative errors calculated for each source and at different fiber lengths are presented in Tab.~\ref{tab:01}.


The experiment is repeated using a Huawei SFP\footnote{The device used is Smartoptics SO-SFP-10GE-LR} with an output power of approximately $-7$ dBm. Fig.~\ref{fig:05} shows the obtained results, while the corresponding relative errors are reported in Tab.~\ref{tab:01}. The good agreement with the data previously observed for the other sources confirms the robustness of the proposed model, demonstrating its applicability to an additional independent source.
Indeed, it is crucial to emphasize that, although this SFP operates within the same C-band wavelength range as the previous sources, its internal architecture and spectral characteristics are inevitably distinct. Any inaccuracies or over-fitting in the model would be immediately exposed by such a change in hardware. Therefore, the consistent performance across different manufacturers and devices proves that the model's validity is not tied to a specific emitter, but rather captures the fundamental physical behaviour of the system, confirming its source-agnostic reliability.
Further confirmation of the model’s robustness comes from measuring the SpRS counts using a spool with a different fiber length. Indeed, the proposed model is applied to all three sources on a 100 m fiber spool. The results are shown in Fig.~\ref{fig:06} and again summarized in Tab.~\ref{tab:01}.
In this last scenario, while the validation of the model remains generally strong, a noticeable deviation from the experimental measurements begins to appear for shorter fiber lengths. Indeed, for fibers of approximately 100 m, Raman counts become less significant compared to those obtained with longer fiber spools.

However, the global trend is still correctly captured, and key features of the response—such as the identification of the minimum—remain reliable and physically consistent. This suggests that measurements with shorter fibers do not significantly compromise the validity of the model, but rather reflect the reduced Raman noise in low-fiber-length scenarios.








\begin{figure}[t]
    \centering
    \includegraphics[width=\linewidth]{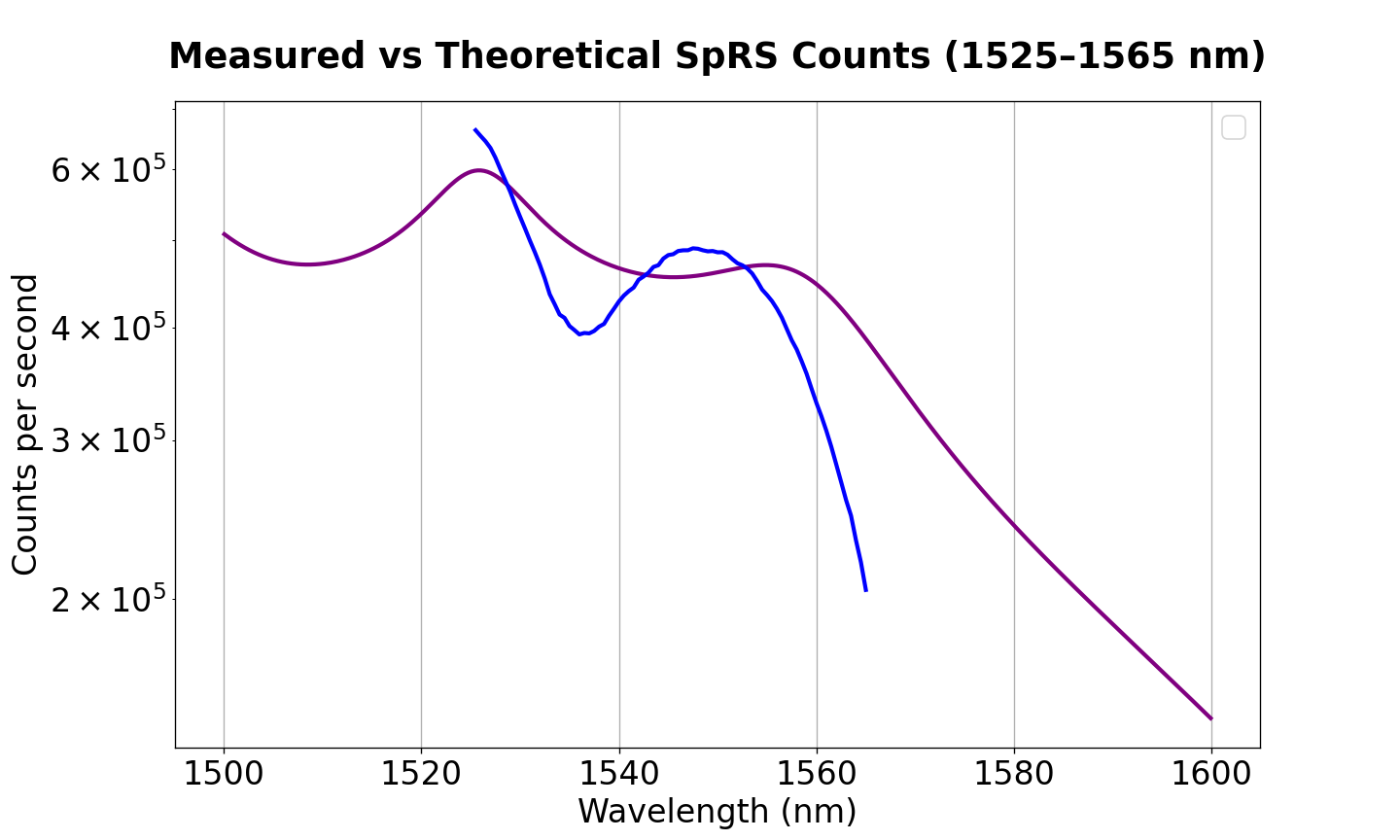}
    \caption{Comparison of SpRS counts: theoretical model versus measured SpRS photons with the laser source at 5000 m.}
    \label{fig:07}
        \hrulefill

\end{figure}

\section{Discussion}
Several theoretical models have already been proposed in the literature to describe and validate Raman scattering in optical fibers \cite{HolCan-02}. These models provide useful insights into the general spectral characteristics of Raman noise generated by a classical pump source. However, these models may not fully capture the complete set of spectral features observed in practical deployment scenarios, particularly under varying source types and urban operating conditions.

In this work, the SpRS counts obtained for different sources and fiber spools of varying lengths exhibit a trend that deviates from the purely theoretical prediction, as shown in Fig.~\ref{fig:07}. In particular, the proposed model—derived directly from experimental data—clearly reveals the presence of a minimum around $\sim 1535$ nm, a feature not reproduced by the theoretical model. Capturing such spectral variations is essential for an accurate description of the Raman noise distribution, which is often oversimplified in analytical approaches.

The resulting fit better reproduces the measured behaviour across the investigated wavelength range and highlights the main spectral features of the Raman profile. These results are further validated by measurements performed in a real urban fiber loop deployment scenario over a 7 km link \cite{CalDavFlo-26}, confirming the robustness of the approach under realistic operating conditions.

Overall, the proposed model enables the prediction of Raman scattering behaviour based on experimentally accessible parameters such as fiber length and source power, without requiring direct spectral measurements for each configuration. This provides a practical tool to estimate the Raman noise profile under different operating conditions. It also helps identify spectral regions where Raman noise is minimized, supporting the optimal selection of channels for quantum signal transmission in hybrid classical–quantum optical networks.

\bibliographystyle{IEEEtran}
\bibliography{bibliography.bib}

@article{CacCalTaf-19,
  title={Quantum internet: Networking challenges in distributed quantum computing},
  author={Cacciapuoti, Angela Sara and Caleffi, Marcello and Tafuri, Francesco and Cataliotti, Francesco Saverio and Gherardini, Stefano and Bianchi, Giuseppe},
  journal={IEEE Network},
  volume={34},
  number={1},
  pages={137--143},
  year={2019},
  publisher={IEEE}
}

@article{CalDavHan-25,
  title={Quantum transduction: Enabling quantum networking},
  author={Caleffi, Marcello and D’Avossa, Laura and Han, Xu and Cacciapuoti, Angela Sara},
  journal={IEEE Communications Surveys \& Tutorials},
  year={2025},
  publisher={IEEE}
}

@article{EraWal-10,
  title={Quantum key distribution and 1 Gbps data encryption over a single fibre},
  author={Eraerds, Patrick and Walenta, Nino and Legr{\'e}, Matthieu and Gisin, Nicolas and Zbinden, Hugo},
  journal={New Journal of Physics},
  volume={12},
  number={6},
  pages={063027},
  year={2010}
}

@article{HolCan-02,
  title={Multiple-vibrational-mode model for fiber-optic Raman gain spectrum and response function},
  author={Hollenbeck, Dawn and Cantrell, Cyrus D},
  journal={Journal of the Optical Society of America B},
  volume={19},
  number={12},
  pages={2886--2892},
  year={2002},
  publisher={Optical Society of America}
}

@inproceedings{ThoKanKum-25,
  title={Multiphoton interference and quantum teleportation coexisting with classical communications in optical fiber},
  author={Thomas, Jordan M and Kanter, Gregory S and Kumar, Prem},
  booktitle={Quantum Computing, Communication, and Simulation V},
  volume={13391},
  pages={98--112},
  year={2025},
  organization={SPIE}
}

@article{FroDynLuc-15,
  title={Quantum secured gigabit optical access networks},
  author={Fr{\"o}hlich, Bernd and Dynes, James F and Lucamarini, Marco and Sharpe, Andrew W and Tam, Simon W-B and Yuan, Zhiliang and Shields, Andrew J},
  journal={Scientific reports},
  volume={5},
  number={1},
  pages={18121},
  year={2015},
  publisher={Nature Publishing Group UK London}
}

@article{CraLazPorSekFlaNam-24,
  title={Automated distribution of polarization-entangled photons using deployed New York City fibers},
  author={Craddock, Alexander N and Lazenby, Anne and Portmann, Gabriel Bello and Sekelsky, Rourke and Flament, Mael and Namazi, Mehdi},
  journal={PRX Quantum},
  volume={5},
  number={3},
  pages={030330},
  year={2024},
  publisher={APS}
}

@article{FerXavTemVon-14,
  title={Impact of Raman scattered noise from multiple telecom channels on fiber-optic quantum key distribution systems},
  author={Ferreira da Silva, Thiago and Xavier, Guilherme B and Tempor{\~a}o, Guilherme P and von der Weid, Jean Pierre},
  journal={Journal of lightwave technology},
  volume={32},
  number={13},
  pages={2332--2339},
  year={2014},
  publisher={OSA}
}

@inproceedings{ClaYanWanNejSimJos-24,
  title={What is the best wavelength for fibre quantum communication?},
  author={Clark, Marcus J and Yang, Ruizhi and Wang, Rui and Nejabati, Reza and Simeonidou, Dimitra and Joshi, Siddarth K},
  booktitle={Quantum 2.0},
  pages={QTh3A--43},
  year={2024},
  organization={Optica Publishing Group}
}

@inproceedings{ThoKanXie-23,
  title={Optimization of classical light wavelengths coexisting with c-band quantum networks for minimal noise impact},
  author={Thomas, Jordan M and Kanter, Gregory S and Xie, Si and Chung, Joaquin and Valivarthi, Raju and Pe{\~n}a, Cristi{\'a}n and Kettimuthu, Rajkumar and Spentzouris, Panagiotis and Spiropulu, Maria and Kumar, Prem},
  booktitle={Optical Fiber Communication Conference},
  pages={Tu3H--3},
  year={2023},
  organization={Optica Publishing Group}
}

@article{FroDynLucSha-15,
  title={Quantum secured gigabit optical access networks},
  author={Fr{\"o}hlich, Bernd and Dynes, James F and Lucamarini, Marco and Sharpe, Andrew W and Tam, Simon W-B and Yuan, Zhiliang and Shields, Andrew J},
  journal={Scientific reports},
  volume={5},
  number={1},
  pages={18121},
  year={2015},
  publisher={Nature Publishing Group UK London}
}

@misc{IlMattino-26,
  author       = {M. Capone},
  title        = {La Quantum valley napoletana: ecco il laboratorio della Federico II per il teletrasporto quantistico},
  howpublished = {Il Mattino},
  year         = {2026},
  month        = jan,
  day          = {17}
}

@article{TalHesDav-26,
  title={Synchronized distribution of quantum entanglement coexisting with high-rate, broadband classical optical communications over a real-world fiber link},
  author={Talcott, Gina M and Hess, Ahnnika I and d'Avossa, Laura and Kohlert, Scott J and Yeh, Fei I and Chen, Jim Hao and Mambretti, Joe J and Rambo, Tim M and Kanter, Gregory S and Thomas, Jordan M and others},
  journal={arXiv preprint arXiv:2602.00253},
  year={2026}
}

@article{CalDavFlo-26,
  author = {Marcello Caleffi and Laura d'Avossa and
            Italo Ignacio Machuca Flores and Marco Grillo and
            Elena Montella and Angela Sara Cacciapuoti},
  title = {Towards Quantum Networks: Characterizing Raman Noise over Metropolitan scale Fiber Network},
  journal = {IEEE SMARTCOMP - Workshop Quantum Computing and Networking for Smart Systems, Invited paper},
  year = {2026}
}

@article{WanRolBri-24,
  title={Time-interleaved C-band Co-propagation of quantum and classical channels},
  author={Wang, Jing and Rollick, Brian J and Jia, Zhensheng and Huberman, Bernardo A},
  journal={Journal of Lightwave Technology},
  volume={42},
  number={11},
  pages={4086--4095},
  year={2024},
  publisher={IEEE}
}

@article{WuRbDiS-25,
  title={Integration of quantum key distribution and high-throughput classical communications in field-deployed multi-core fibers},
  author={Wu, Qi and Ribezzo, Domenico and Di Sciullo, Giammarco and Cocchi, Sebastiano and Ann Shaji, Divya and Alves Zischler, Lucas and Luis, Ruben and Serena, Paolo and Lasagni, Chiara and Bononi, Alberto and others},
  journal={Light: Science \& Applications},
  volume={14},
  number={1},
  pages={274},
  year={2025},
  publisher={Nature Publishing Group UK London}
}

@article{DavCacCal-25,
  title={Modeling quantum transduction for multipartite entanglement distribution},
  author={d'Avossa, Laura and Cacciapuoti, Angela Sara and Caleffi, Marcello},
  journal={IEEE Transactions on Communications},
  volume={73},
  number={11},
  pages={11707--11721},
  year={2025},
  publisher={IEEE}
}

@inproceedings{TalHesTho-26,
  title={Synchronized Entanglement Distribution Across Deployed Fiber Coexisting with Fully-Loaded C-band Optical Communications},
  author={Talcott, Gina M and Hess, Ahnnika I and Thomas, Jordan M and d’Avossa, Laura and Kohlert, Scott J and Yeh, Fei I and Chen, Jim Hao and Mambretti, Joe J and Rambo, Tim M and Kanter, Gregory S and others},
  booktitle={Optical Fiber Communication Conference},
  pages={Tu2K--2},
  year={2026},
  organization={Optica Publishing Group}
}

\end{document}